\documentclass[12pt]{article}
 \usepackage[dvips]{graphicx}
\begin{document}
\title {Extended Scale Relativity, p-Loop Harmonic Oscillator and Logarithmic Corrections to the Black Hole
Entropy}

\renewcommand{\baselinestretch}{1}
\bigskip
\author{Carlos Castro \thanks{Center for Theoretical Studies of Physical
Systems,Clark Atlanta University,Atlanta, GA.
30314;~E-mail:castro@ts.infn.it}~~ and ~ Alex
Granik\thanks{Department of Physics, University of the Pacific,
Stockton,CA.95211;~E-mail:agranik@uop.edu}}
\date{}
\maketitle \vspace{10cm} \pagebreak
\renewcommand{\baselinestretch}{1.6}
\begin{abstract}
 A new relativity theory, or more concretely an extended
relativity theory, actively developed by one of the authors
incorporates 3 basic concepts. They are the old idea of Chew
about bootstrapping, Nottale's scale relativity, and enlargement
of the conventional time-space by inclusion of noncommutative
Clifford manifolds where all p-branes are treated on equal
footing. The latter allows one to write a master action
functional. The resulting functional  equation is simplified and
applied to the p-loop oscillator. Its solution represents a
generalization of the conventional quantum point oscillator. In
addition , it exhibits some novel features:  an emergence of two
explicit scales delineating the asymptotic regimes (Planck scale
region and a smooth region of a quantum point oscillator). In the
most interesting Planck scale regime, the solution recovers in an
elementary fashion some basic relations of string theory
(including string tension quantization and string uncertainty
relation). It is shown that the degeneracy of the $first$
collective excited state of the  p-loop oscillator yields not
only the well-known Bekenstein-Hawking area-entropy linear
relation but also the logarithmic corrections therein. In
addition we obtain for any number of dimensions the Hawking
temperature, the Schwarschild radius, and the inequalities
governing the area of a black hole formed in a fusion of two
black holes. One of the interesting results is a demonstration
that the evaporation of a black hole is limited by the upper
bound on its temperature, the Planck temperature.
\end{abstract}

\section{The $p$-Loop Harmonic Oscillator in Clifford Manifolds }
\renewcommand{\baselinestretch}{2}
An idea that condensed-matter-like quantum theories of gravity
might be the reason behind general relativity was know  for quite
some time going back to A.Sakharov \cite{as1},\cite{wu}. Quite
recently, G.Chapline, R.Laughlin {\it et al} proposed $\it
collective$ effects  as an explanation for the apparent
incompatibility of quantum mechanics and general relativity
\cite{gc1}. It was also argued \cite {gc2} that superstring
models can be only an approximation to a more fundamental theory.
A finite quantum model for the ground state of 4-dim gravity
based on a condensate wavefunction was proposed \cite{gc3}as an
attractive alternative to superstring theory. Once again the
$collective$ effects play a crucial role in formulating the
theory. There is also an emerging consensus that basic physical
constants such as $\hbar$, speed of light $c$ might be
scale-dependent where their deviation from constant values
becomes essential at scales comparable with the Planck length
$\Lambda \sim 10^{-35}m$.

All these ideas gave impetus to combining them under one "roof"
of what could be loosely called extended relativity.
Interestingly enough, one of the ways to describe collective
effects is to use Clifford multivectors which allow one to treat
objects of any dimensionality ( scalars, vectors, multivectors)on
equal footing. Extended relativity \cite{cc1}-\cite{cc10} has
introduced a generalization of the ordinary space-time  by
considering non-commutative spaces where all $p$-branes ($p=
0,1,...n$) are unified on the basis of Clifford multivectors
(prompting one to call such spaces C-spaces, that is spaces
spanned by Clifford multivectors). This resulted in a fully
covariant $p$-brane quantum mechanical loop wave equations which
follow from a master action functional that has been motivated by
earlier work \cite{cc3},\cite{aa1}.

Such master action functional $S\{\Psi [X(\Sigma)]\}$ of quantum
field theory in C-space \cite{cc8}-\cite{cc4} has the following
form
\begin{equation}
\begin{array}{c}
S\{\Psi [X(\Sigma)]|Q,P]\}= \int~ [DX(\Sigma)]DPDQ ~ [{1\over 2}
~({ \delta \Psi \over
 \delta X }*{ \delta \Psi \over \delta X }+ {\cal E}^2 \Psi*\Psi)+
 {g_3\over 3!} \Psi*\Psi*\Psi +\\
 {g_4\over 4!}
 \Psi*\Psi*\Psi*\Psi].
\end{array}
\label{eq:o1}
\end{equation}

Here $\Psi [X(\Sigma)]$ is the Clifford scalar field , $\Sigma$ is
an invariant evolution parameter (a generalization of the proper
time in special relativity) such that
\begin{equation}
\begin{array}{l}
  (d\Sigma)^2 = (d\Omega_{p+1})^2 + \Lambda^{2p}(dx_\mu dx^\mu)+
  \Lambda^{2(p-1)}d\sigma_{\mu\nu}d\sigma^{\mu\nu}+...\\
  +(d\sigma_{\mu_1\mu_2...\mu_{p+1}}d\sigma^{\mu_1\mu_2...\mu_{p+1}}),
  \end{array}\label{eq:o2}
\end{equation}

\begin{equation}
\label{eq:o3}
 X(\Sigma)=\Omega_{p+1}I+\Lambda^p x_\mu \gamma^\mu+
 \Lambda^{p-1}\sigma_{\mu\nu}\gamma^\mu\gamma^\nu+...
\end{equation}

is a Clifford multivector "living" on the Clifford manifold
outside space-time, $\Lambda$ is the Planck scale that allows one
to combine objects of different dimensionality in
Eqs.(\ref{eq:o2},\ref{eq:o3})(serving as the $universal$
reference length). The multivector $X$ Eq.(\ref{eq:o3}) combines
both a point history given by the ordinary (vector) coordinates
$x_\mu$ and the holographic projections of the nested family of
all $p$-loop histories onto the embedding coordinate spacetime
hyperplanes $\sigma_{\mu\nu},...\sigma_{\mu_1 \mu_2...\mu_{p+1}}$.

The scalar ( from the point of view of ordinary Lorentz
transformation but $not$ from the C-space point of view )
$\Omega_{p+1}$ is the invariant proper $p+1=D$-volume associated
with a motion of a maximum dimension $p$-loop across the $p+1
=D$-dim target spacetime. The multivector $X$ has $2^D$
components.

Variables $$Q=\{q_i, q_{ij},
q_{ijk},...\}~~P=\{p_i,p_{ij},p_{ijk},...\}$$ are multivectors
representing "coordinates" and "momenta" in C-space and

\begin{center}
 $  DQ=dq_{i}dq_{ik}...,~~~
  DP=dp_{\i}dp_{ik}...$,\\
  \end{center} Finally, the
star product "$*$" in Eq.(\ref{eq:o1}) is an extension to the
C-space of the Baker-Vasiliev star product \cite{b},\cite{MV}
\begin{equation}
\label{eq:o4}
 (F*G)(Q,P)=({1 \over 2\pi\hbar})^{2r} \int
 DQ'DP'DQ''DP''e^{iK}F(Q',P')G(Q'',P'')
\end{equation}
where
\begin{center}
$r=D + {D(D-1)\over 1!}+ {D(D-1)(D-2)\over 2!}+ {D(D-1)(D-2)(D-3)
\over 3!}+...=2^{D-1}D$\\
\end{center}

the kernel $$K =Z_A\Omega^{AB}Z_B=
(Q',P'')-(Q'',P')+(Q'',P)-(Q,P'')+(Q,P')-(Q',P),$$
 and $(,)$ represents a dimensionless ( normalized by $\hbar$) scalar part of the
 Clifford product
\begin{center}
$(Q,P')={q_{\mu}p^{'\mu}\over{\hbar}}+{q_{\mu\nu}p^{'\mu\nu}\over{\hbar^2}}+...$
\end{center}
\smallskip

As an aside note, we have to mention that the "classical" field
theory in C-spaces requires the use of a non-commutative but still
associative Baker-Vasiliev star product reflecting a slight
modification of the conventional Moyal product \cite{JM}  .
However quantization of the master action (\ref{eq:o1}) would
require an introduction of a more "sophisticated" q-star product
which is not associative any more and represents a quantum group
deformation of the Moyal product related to a non-associative
geometry and quantum Clifford algebras
{\cite{AN},\cite{BF},\cite{ZO},\cite{OD}.

In what follows we concentrate our attention on a truncated
version of the theory by applying it to a linear $p$-loop
oscillator.

Such a  truncated ( linearized) version follows from $3$
simplifications. First, in (\ref{eq:o1}) the cubic and the
quartic terms are dropped. Secondly, all the holographic modes
are frozen  and only the zero modes are kept. This would yield
conventional "master" action instead of the functional one
generated with the help of the Moyal product \cite{OD},
non-associative geometry {\cite{AN}, and quantum Hopf algebra
\cite{BF}, \cite{ZO}. Finally, we assume that the metric in
C-space is flat.\\

Now we define what  "relativistic" means in terms of the extended
relativity. The complete theory is the master field $\Psi
[X(\Sigma)]$ theory whose action functional admits a
noncommutative braided quantum Clifford algebra. As a result of
the postulated simplifications, we are performing a reduction of
such a field theory to an ordinary quantum mechanical theory.
Still we have to keep in mind that fields are $not$ quantized wave
functions. For this reason the wave equations that will emerge
describe a nonrelativistic theory in {\bf C}-spaces.

Applying the above simplifying assumptions to the action
(\ref{eq:o1})  we obtain the following wave equation describing a
 p-loop linear oscillator in C-space
\begin{equation}
\begin{array}{l}
  \{-{1\over2}{1\over\Lambda^{p-1}}~[\frac{\partial^2}{\partial
 {x_\mu}^2}+\Lambda^2\frac{\partial^2}{(\partial\sigma_{\mu\nu})^2}+
 \Lambda^4\frac{\partial^2}{(\partial \sigma_{\mu\nu\rho})^2}+...+
 \Lambda^{2p}\frac{\partial^2}{(\partial\Omega_{p+1})^2}]+\\

 {m_{p+1}\over2}{1\over L^2}~[\Lambda^{2p}x{_\mu}^2+\Lambda^{2p-2}
{\sigma_{\mu\nu}}^2+...+\Omega^2_{p+1}]\}~\Psi = T\Psi
 \end{array} \label{eq:o5}
\end{equation}

In Eq.(\ref{eq:o5}) $T$ is tension of the spacetime-filling
$p$-brane related to the "mass-parameter-like" quantity $\cal E$
(present in the action functional Eq.(\ref{eq:o1}))via
\begin{center}
${\cal E}^2 = m_{p+1}T + (m^2_{p+1}V_{harmonic}/L^2);$
\end{center}
where $V_{harmonic}$ is the harmonic potential for the $p$-loop
oscillator. $ D = p+1$, $m_{p+1}$ is the parameter of dimension
$(mass)^{p+1}$ , the parameter $L$ (to be defined later) has
dimension $(length)^{p+1}$ and we use units $\hbar = 1, c = 1$.

For the following analysis we rewrite Eq.(\ref{eq:o5}) in  the
dimensionless form
\begin{equation}
 \{ \frac{\partial^2}{\partial \tilde{x}_{\mu}^2}+
\frac{\partial^2}{\partial \tilde{\sigma}_{\mu\nu}^2}+...-
(\tilde{\Omega}^2+\tilde{x}_{\mu}^2+\tilde{\sigma}_{\mu\nu}^2+...)+
 2{\cal{T}}\}\Psi =0
 \label{eq:o6}
\end{equation}
where ${\cal{T}}= \frac{T}{ m_{p+1}\sqrt{\cal A}}$ is the
dimensionless tension.  $\cal A$ is a scaling parameter that will
be determined below.
\begin{equation}
   \tilde {x}_{\mu}\hspace{1mm} =
  {\cal{A}}^{1/4}\frac{\Lambda^p}{L}x_\mu,\hspace{3mm}
    \tilde{\sigma}_{\mu\nu}\hspace{1mm} = {\cal{A}}^{1/4}\sigma_{\mu\nu}\frac{\Lambda^{p-1}}{L},...,
\tilde{\Omega}_{p+1}={\cal{A}}^{1/4}\frac{\Omega_{p+1}}{L}
\label{eq:o7}
\end{equation}
are the dimensionless arguments such that  $\tilde{x}_{\mu}$ has
$D$ components, $\tilde{\sigma}_{\mu\nu}$ has $ [D!]/[(D-2)!2!]$
components, etc.

Inserting the usual Gaussian solution for the ground state into
the wave equation Eq.(\ref{eq:o6}) we obtain the value of ${\cal
A}$ :
\begin{equation}
 {\cal{A}} \equiv {m_{p+1}L^2}/{\Lambda^{p+1}}
 \label{eq:o8}
\end{equation}
Without any loss of generality we can set ${\cal{A}}=1$ by
absorbing it into $L$. This will give the following geometric
mean relation between the parameters $L, \hspace{1.5mm}m_{p+1}$,
and $\Lambda$

$$L^2 = {\Lambda^{p+1}}/{m_{p+1}} \Rightarrow \Lambda^{p+1} < L
<   { 1 \over m_{p+1} }   $$ indicating a presence of three
scaling regimes. Their respective meaning is as follows.

1) The scale given by generalized Compton wavelength $  (1/
m_{p+1})^{( 1/p+1) }$ signals a transition from a smooth
continuum to a $fractal$ ( but continuous ) geometry.

2)The scale $L$ determines a $discrete$ and a $ fractal$ world,
which is similar to the ones found in El Naschie's
Cantorian-fractal spacetime models \cite{meln} and $p$-adic
quantum mechanics \cite{matpitka}

3)Finally $\Lambda$ corresponds to the quantum gravitational
regime.

Using ${\cal A} =1$ we get from Eq.(\ref{eq:o7}) $$\tilde{x}_\mu =
\sqrt{\Lambda^{p+1}m_{p+1}}\hspace{1mm}\frac{x_\mu}{\Lambda},~~
\tilde{\sigma}_{\mu\nu}
=\sqrt{\Lambda^{p+1}m_{p+1}}\hspace{1mm}\frac{\sigma_{\mu\nu}}{\Lambda^2},...,
\tilde{\Omega}_{p+1}=\sqrt{\Lambda^{p+1}m_{p+1}}\hspace{1mm}
\frac{\Omega_{p+1}}{\Lambda^{p+1}}\eqno(8a)$$

\bigskip
The dimensionless combination $\Lambda^{p+1}m_{p+1}$ (which
indicates existence of two separate scales , $\Lambda$ and
$(1/{m_{p+1}})^{\frac{1}{p+1}}$) obeys the following double
inequality:
\begin{equation}
\label{eq:9}
\ \sqrt{{m_{p+1}\Lambda^{p+1}}} < 1 <
\sqrt{\frac{1}{m_{p+1}\Lambda^{p+1}}}
\end{equation}
In turn, relations (\ref{eq:9}) define two asymptotic regions:

1)the "discrete-fractal" region characterized by
${m_{p+1}\Lambda^{p+1}} \sim 1$, or the Planck scale regime, and

\smallskip
2)the "fractal/smooth phase transition ", or the low energy region
characterized by ${m_{p+1}\Lambda^{p+1}}<< 1$.

\bigskip
Since the wave equation (\ref{eq:o6}) is diagonal in its arguments
(that is separable) we represent its solution  as a product of
separate functions of each of the dimensionless arguments
$\tilde{x}_{\mu},\tilde{\sigma}_{\mu\nu},etc. $
\begin{equation}
\label{eq:10}
  \Psi =\prod _{i}F_i(\tilde
{x}_i)\prod_{j<k}F_{jk}(\tilde\sigma_{jk})...
\end{equation}

Inserting (\ref{eq:10}) into (\ref{eq:o6}) we get for each of
these functions the Whittaker equation:
\begin{equation}
\label{eq:11}
    Z^{\prime\prime} - (2{\cal{T}} - \tilde {y}^{2})Z =0
\end{equation}
where $Z$ is any function $F_i, F_{ij},...$, $\tilde{y}$ is the
respective dimensionless variable $\tilde{x}_{\mu}$,$\tilde
{\sigma}_{\mu\nu},...$, and there  are all in all $2^D$ such
equations. The bounded solution of (\ref{eq:11}) is expressed in
terms of the  Hermite polynomials $H_n(\tilde{y})$
\begin{equation}
\label{eq:12}
    Z \sim e^{-\tilde{y}^2/2}H_n(\tilde{y})
        \end{equation}
Therefore the solution to Eq.({\ref{eq:o6}) is
\begin{equation}
\label{eq:13}
 \Psi \sim
 exp[-({\tilde{x}_\mu}^2+{\tilde{\sigma}_{\mu\nu}}^2+...+\tilde{\Omega}^2_{p+1})]
 \prod_{i}H_{n_i}(\tilde{x}_i)\prod _{jk}H_{n_{jk} }(\tilde{\sigma}_{jk})...
 \end{equation}
where there are $D$ terms corresponding to $n_{1},n_{2},...,
n_{D}$. There are $D(D-1)/2$ terms corresponding to holographic
area excitation modes $n_{01}, n_{02},...,$ etc. Thus the total
number of terms corresponding to the $N$-th excited state
$(N=n_{x1}+n_{x2}+...+n_{\sigma_{01}}+n_{\sigma_{02}}+...)$  is
given by the degree of the Clifford algebra in $D$ dimensions,
that is $2^D$.

The respective value of the tension of the $N$-th excited state is
\begin{equation}
\label{eq:14}
 T_N= (N + \frac{1}{2} 2^D) m_{p+1}
\end{equation}
yielding  {\bf quantization}  of tension.

Expression (\ref{eq:14}) is the analog of the respective value of
the $N$-th energy state for a quantum point oscillator. The
analogy however is not complete. We point out one substantial
difference. Since according to a new relativity principle
\cite{cc1}, \cite{cc8} all the dimensions are treated on equal
footing (there are no preferred dimensions) all the modes of the
$p$-loop oscillator( center of mass $x_\mu$, holographic modes,
$p+1$ volume) are to be excited collectively. This behavior is in
full compliance both with condensed-matter-like quantum theories
of gravity (mentioned earlier ) and with the principle of
polydimensional invariance introduced by Pezzaglia \cite{wp1}
postulating that physical laws should be invariant under  local
automorphism transformations "reshuffling" the physical geometry.
In particular, this means that vectors, bivectors ( areas),
trivectors, etc. all viewed as  parts of one multivector, and as
such can be transformed into one another by the local
automorphism transformations.

Based on this considerations, the first excited state is not $N=1$
( as could be naively expected) but rather $N=2^D$. Therefore
\begin{center}
    $T_1 \rightarrow T_{2^D} = \frac{3}{2}(2^D m_{p+1}$)
\end{center}
instead of the familiar $(3/2)m$.

Having obtained the solution to Eq.(\ref{eq:o6}), we consider in
more detail the two limiting cases corresponding to the above
defined 1) fractal and 2) smooth regions. The latter (according
to the correspondence principle) should be described by the
expressions for a quantum point oscillator. In particular, this
means that the analog of the zero slope limit of string theory
(the field theory limit) represents a collapse of the $p$-loop
histories to a point history :

$$ \Lambda \rightarrow 0,~~ m_{p+1} \rightarrow \infty,~~ T
\rightarrow \infty,~~ \sigma_{\mu\nu},
\sigma_{\mu\nu\rho},...      \rightarrow   0,~~~ L \rightarrow
0.~ $$.

These limits are taken in such a way that the following
combination reproduces the standard results of a point-particle
quantum oscillator :

\begin{equation}
\label{eq:15}
 \tilde{x}_{\mu}=
\frac{x_{\mu}}{\Lambda}\sqrt{{m_{p+1}\Lambda^{p+1}}} \rightarrow
x_{\mu}/a
\end{equation}
where the $nonzero$  parameter $a > \Lambda$ is a finite
quantity, namely the amplitude of the conventional point-particle
oscillator.

On the other hand, in string theory, there are two scales, the
Planck scale $\Lambda$ and the string scale $ l_s > \Lambda$.
Without any loss of generality we can set $ a \sim l_s$. A large
value of $a >> \Lambda$ would then correspond to a ``macroscopic''
string. We shall return to this point when we address the
black-hole entropy.

Using Eq.(\ref{eq:15}) we find ${m_{p+1}}$ in terms of the other
variables :
 \begin{equation}
 \label{eq:16}
 {m_{p+1}}\rightarrow{(M_{Planck})}^{p+1}
 (\frac{\Lambda}{a})^2 < {(M_{Planck})}^{p+1}
 \end{equation}
where the Planck mass $M_{Planck} \equiv 1/\Lambda$. Notice that
in the field-theory limit, $\Lambda \rightarrow 0$, when the loop
histories collapse to a point-history, (\ref{eq:16}) yields $
{m_{p+1}} \rightarrow \infty$ as could be expected. From
Eqs.({\ref{eq:14}) and (\ref{eq:16}) follows that in this region
\begin{equation}
\label{eq:17}
 \ {T_N} \sim
 (M_{Planck})^{p+1}(\frac{\Lambda}{a})^2< (M_{Planck})^{p+1}\\
\end{equation}
in full agreement with  this region's scales as compared to the
Planck scales.

At the other end of the spectrum ( discrete-fractal/quantum
gravity  region) where
 $m_{p+1}\Lambda^{p+1}\sim 1$ we would witness a collapse of all
the scales to only one scale, namely the Planck scale $\Lambda$.
In particular, this means that the string scale $a \sim l_s \sim
\Lambda$, and the oscillator parameters become
\begin{equation}
\label{eq:18}
 \ \tilde{x}_{\mu}= \frac{x_{\mu}}{\Lambda}
\sqrt{\Lambda^{p+1}m_{p+1}} \sim
\frac{x_\mu}{\Lambda},\hspace{2mm} m_{p+1} \sim
\frac{1}{\Lambda^{p+1}} \equiv (M_{Planck})^{p+1},
\end{equation}

The ground state tension is :

$$T_o \sim m_{p+1} \sim\frac{1}{\Lambda^{p+1}}$$ These relations
are the familiar relations of string theory. In particular, if we
set $p=1$ we get the basic string relation
\begin{center}
$ T \sim \frac{1}{\Lambda^2} \equiv \frac{1}{\alpha'}$
\end{center}

\smallskip
Above we got two asymptotic expression for $m_{p+1}$
\[ m_{p+1} = \left\{\begin{array}{ll}
\ \Lambda^{-(p+1)}(\Lambda/a)^2 & \mbox{if~
$\Lambda/a < 1$}\\
\ \Lambda^{-(p+1)} & \mbox{if ~$m_{p+1}\Lambda^{p+1}\sim 1,~a
\sim \Lambda$}
  \end{array}
  \right. \]
It is suggestive to write $m_{p+1}\Lambda^{p+1}$ as power series
in $(\Lambda/a)^2$ (cf. analogous procedure in hydrodynamics
\cite{DV1}):
\begin{equation}
\label{eq:19}
 m_{p+1}\Lambda^{p+1}\equiv F(\Lambda/a)
=(\frac{\Lambda}{a})^2[1+ \alpha_1
(\frac{\Lambda}{a})^2+\alpha_2(\frac{\Lambda}{a})^4+...]
\end{equation}
where the $small $ coefficients  $\alpha_i$ are such that the
series is convergent for $a \sim \Lambda$.

Using Eq.(\ref{eq:19}), we then obtain that in the field theory
limit $\Lambda \rightarrow 0$ the dimensionless coordinate
${\tilde x}_\mu$ given by Eq.(8a), becomes :

$${\tilde x}_\mu = {x_\mu \over a} [ 1 + {\alpha_1 \over 2}
({\Lambda \over a})^2 +...] \rightarrow  {x_\mu \over a}$$ where
we use expansion of the square root in Taylor series. Notice that
$a$ is a finite nonzero quantity.

\smallskip
If $p=1 ~(p+1 = D = 2) $ then  for the ground state $N =0$ Eq.
(\ref{eq:14}) yields the ground energy per unit string length :
$T_{ground}=2m^2$ (see footnote\footnote{that is for a point
oscillator we get in units of $\hbar$ and $c$
$E_{ground}=\hbar\omega/2=\sqrt{T_{ground}/8}$}). Returning to
the units $\hbar , c$, introducing $c/a =\omega$ ( where $\omega$
is the characteristic frequency and $a$  is now treated as a
characteristic wavelength $\lambda$), and using (\ref{eq:19}) we
get (cf.ref \cite{cc5})
\begin{center}
$\hbar_{eff} = \hbar \sqrt{1+ \alpha_1
(\frac{\Lambda}{a})^2+\alpha_2(\frac{\Lambda}{a})^4+...}$
\end{center}
Truncating this series at the second term, and using the relations
$h/\lambda \equiv\hbar k =\langle \sqrt{|p|^2}\rangle >
\sqrt{{|\Delta p}|^2}$  we recover the well-known string
uncertainty relation\footnote {here $p$ is the momentum, not to
be confused with our notation used for a $p$-loop oscillator,
$\langle...\rangle$ denotes quantum averaging} \cite{cc5}:
\begin{center}
$ \Delta x \Delta p >  | [ x , p ]|  = \hbar_{eff}  \Rightarrow
\Delta x >  {\hbar \over\Delta p} + \beta {\Lambda^2 \over \hbar
} \Delta p $ \end{center} where $\beta$ is a multiplicative
parameter and $[,]$ is the usual commutator. As $\Lambda
\rightarrow 0$ one recovers the ordinary Heisenberg uncertainty
relations. Interestingly enough, on one hand the string
uncertainty relation until recently did not have " a proper
theoretical framework for the extra term" \cite{ew1}. On the
other hand, this relation has emerged as one of the results of
an  extended relativity in general  \cite{cc5} and our model of a
p-loop oscillator in particular.

\smallskip
\section{A Self Gravitating gas of $p$-loop oscillators and the Schwarzschild Radius
Relation}

So far we have been discussing the collective excitations of a
$p$-loop oscillator without addressing a problem of how such
collective excitations contribute to an emergence of the black
hole geometry.  There exists a related, but more profound
question about emergence of the Riemannian geometry as a long
range limit of $C$-space geometry (where the $p$-loop oscillator
"resides"). However despite of its importance this problem lies
outside the scope of the present work.

Here we argue ( a more detailed an rigorous discussion is given
in \cite{mp}) that the scalar curvature in a C-space can be
expanded in powers of the ordinary Riemannian curvature and its
derivatives. The latter (higher derivatives) are associated with
the higher derivative gravity. A hidden origin of curvature (not
explicitly present in the equations of section $2$, e.g.
Eq.(\ref{eq:o6}) is contained ( or encoded) in the derivatives
with respect to the holographic coordinates, for example, in
$\partial \over {\partial \sigma_{\mu\nu}}$. The units of this
(inverse length squared) coincide with the ones for the curvature.
A more detailed analysis reveals that a long range limit of a
C-space recovers Riemannian geometry and in particular
Riemann-Cartan spaces characterized not only by curvature but
also by torsion \cite{mp}.\\

Therefore we treat ( cf. \cite{gc2}) an emergence of the black
hole geometry as a result of a condensation of the collective
quanta excitations of the p-loop oscillators, keeping in mind the
emergence of the curvature is ensured by the presence of the
holographic coordinates and their derivatives. We consider a
self-gravitating spherically symmetric gas of $p$-loop
oscillators (spherical oscillating "bubbles") whose overall size
and mass we would identify as the Schwarzschild radius $R$ and
the black hole mass $M$ respectively.

Using an analogy between ordinary point and a $p$-loop
oscillators we write
\begin{equation}
\label{eq:20}
 Tension=T \sim m_{p+1}\Omega^2 L^2
\end{equation},
  where $\Omega$
and $L$ play the roles of "frequency" and "amplitude"
respectively. Inserting tension quantization condition Eq.
(\ref{eq:14}) into (\ref{eq:20}) we get the $p$-loop
amplitude-frequency quantization condition
\begin{equation}
\label{eq:21} \ \Omega L = ( N + { 2^D\over2})^{1/2}
 \end{equation}

As we have already discussed, the $first$ collective excitation
corresponds to $ N = 2^D$. Therefore for this case the
quantization condition (Eq.\ref{eq:21}), yields
\begin{equation}
\label{eq:22}
 \ \Omega L ~\sim  N^{1/2} = 2^{D/2} \rightarrow
(\Omega L)^{1/D} \sim 1
\end{equation}

We will view the black hole mass and radius as true zero-point
fluctuations of the gas of $p$-loop oscillators. Therefore with
the help of (\ref{eq:22})  we can represent its mass and the
radius in terms of the characteristic "amplitude" $\lambda \equiv
L^{1/D}$ and "frequency" $\overline{\omega} \equiv \Omega^{1/D}$
\begin{equation}
\label{eq:23}
 M = N^{1/2} \overline{\omega};~~~ R = N^{1/2}\lambda
 \end{equation}
These relations look like relations found in random walk models.
From Eq. (\ref{eq:23}) we immediately obtain the Regge relation
\begin{equation}
\label{eq:24}
 MR = N
 \end{equation}

The geometric mean relation found earlier $$ \Lambda^D<L< {1\over
m_{p+1}}$$ becomes after raising each term to the $1/D$ power $$
\Lambda < L^{1/D} \equiv \lambda < ({ 1\over m _{p+1}})^{1/D}.$$
Using Eq.(\ref{eq:23})and the fact that $\overline{\omega}
\lambda \sim 1$ we get the following identity $$N= MR= ({R\over
\lambda})^2.$$ We assume ( in agreement with the results of
\cite{l1})that a number of bits in a  $(d-2)$-dimensional horizon
area in Planck units is given by $$ N = (R/\Lambda)^{ d-2}$$
(later we discuss this relation in more detail). Upon substitution
of the above identity into the number of bits , we arrive at an
important relation
\begin{equation}
\label{eq:25} ({\lambda \over\Lambda})^{2 \over{4-d}}={R
\over\Lambda}
\end{equation}

Since number of bits $N \ge 1$ and $\Lambda$ is the minimum
attainable length scale, Eq.(\ref{eq:25}) restricts the allowed
range of the values of $d$:
\begin{equation}
\label{eq:26} 4
 \ge d \ge 2
\end{equation}
For $d=4$ the exponent becomes singular yielding $\lambda
=\Lambda$. The value $d=3$ results in  the following mean
geometric relation $$\lambda^2 =R\Lambda \rightarrow \Lambda <
\lambda <R.$$ Finally, for $d=2$ $$\lambda =R$$ which means that
$N =1.$ Since Clifford algebras are defined for $D \ge 2$ ( that
is $N_{min}=2^2=4 >1$) we arrive at a conclusion that for any
$N>1$ the dimensions $d$ must fall into the range which is the
modification of Eq.(\ref {eq:26}): $$ 2 < d \leq 4.$$

We would like to emphasize that we are considering a "toy" model
of a black hole, and therefore it is not surprising that the
number of dimensions $d$ is rather restricted. On the other hand,
the fact that $d=4$( physical spacetime) describes a special case
of $\lambda =\Lambda$ ( the characteristic length equals the basic
length scale, Planck scale), and  $d \ge 2$ tells us that even
within the restrictions necessarily inherent to the "toy" model
we obtain physically sound results. This speaks in favor of our
approach.

As a next step we derive the Scwarzschild radius-mass relation,
once again using a spherically-symmetric self-gravitating gas of
quanta associated with a set of $p$-loop oscillators. Let us
consider such a gas in spacetime of $d$ dimensions. Its
"characteristic" pressure ( or energy density) at the surface of a
radius $r=R$ is
\begin{equation}
\label{eq:27}
 p \sim {E \over {R^{d-1}}}= {M \over{R^{d-1}}}
\end{equation}
where $M$ is the total mass of the gas. To write the exact
relation , one must know the exact equation of state for such a
gas ( see for example, \cite{gc2}). Meanwhile we use further
simplification assuming that the pressure attains its maximum
 at the center $r=0$ and reaches its minimum value at the surface
$r=R$ of the gas.

On the basis of dimensional considerations we arrive at a
conclusion that the pressure counterbalancing gravitational
attraction for such a gas has the same units as the $p$-brane
tension $T$ (energy density) for the special case of a spacetime
filling $p$-brane (here $p$ is the dimension of the brane, $p
=d-1$). Therefore for the first collectively excited state
introduced earlier we obtain with the help of the bit-area
relation $N = (R/\Lambda)^{ d-2}$ the following
\begin{equation}
\label{eq:28}
p \sim T\sim {N\over {R^d}} \sim({R
\over\Lambda})^{d-2}{1 \over R^d}
\end{equation}

Using Eqs.(\ref{eq:27})-(\ref{eq:28}) we obtain that the pressure
$p$ (not to be confused with a notation $p$ used in $p$-brane) is

\begin{equation}
\label{eq:29}
 ({R \over \Lambda })^{d-2}{1 \over R^d} \sim {M
\over R^{d-1}} \rightarrow M \sim {R^{d-3}\over \Lambda^{d-2}}
\rightarrow  R \sim (\Lambda^{d-2} M)^{1/(d-3)}.
\end{equation}
On the other hand, the Newtonian constant $G_d$ in $d$ dimensions
is
\begin{equation}
\label{eq:30}
 G_d = \Lambda^{d-2}
\end{equation}
As a result, we get the Schwarschild radius-mass relation for a
black hole in $d$ dimensions with a respective $d-2$-dimensional
area: $$R \sim (G_d M)^{1/(d-3)}.$$ This means that we can
interpret the Schwarschild radius $R$ as the maximum radius of a
stable spherically symmetric gas of "quanta" associated with the
first collectively excited state of a $p$-loop oscillator with a
characteristic mass $M$.

Our approach also allows one to arrive at the Hawking formula for
the temperature ${\bf T}$ of a black hole. Using
 equation of state for a perfect gas $pV=Nk {\bf
T}$ (remembering that now the volume is proportional to
$R^{d-1}$), the Regge relation $MR =N,$ and setting the Boltzmann
constant $k =1$ we get
\begin{equation}
\label{eq:31}
p V \sim p R^{d-1}\sim M \sim N{\bf T}=MR{\bf T}
\rightarrow R \sim{1 \over {\bf T}}
\end{equation}
where $p$ is the pressure, the number $N$ plays the role of the
number of "molecules" of the gas and we use Eq.(\ref{eq:27}).

Upon substitution of the Schwarschild mass-radius relation into
(\ref{eq:31})  we obtain $$GM^{1/(d-3)} \sim ({1 \over {\bf T}})$$
which is the generalization of the famous Hawking formula $GM
\sim {1 \over {\bf T}}$  ( corresponding to $d=4$) to an
arbitrary number of dimensions $d$ (in fact, as we have shown
earlier $2 < d \leq 4$). More precise relation between the mass
and the temperature can be derived by using the thermodynamic
identity (in units where the Boltzmann constant $k =1$) $${1
\over {\bf T}}={\partial S \over {\partial E}}$$ where the
entropy $S$ as a function of the black hole area will be found in
Section $3$.

\section{Black Hole Entropy as a Function of the Area}
We apply the obtained results to the determination of the black
hole entropy. To this end we find the degeneracy associated with
the $N$-th excited level of the $p$-loop oscillator. The
degeneracy $dg(N)$ is equal to the number of partitions of the
number $N$ into a set of \hspace{1mm}  $2^D$ numbers $$N =
\{n_{x1} + n_{x2}+ ...+ n_{xD}+ n_{\sigma_{\mu\nu}}+
n_{\sigma_{\mu\nu\rho}}+ ...+ n_{ \Omega_{p+1} } \}.$$ Once again
we see the emergence of $collective$ center of mass excitations,
holographic area excitations , holographic volume
excitations,etc. They are given by the set of quantum numbers
$n_{x_D};n_{\sigma_{\mu\nu}},..., n_{ \Omega_{ p+1} } $
respectively. These collective $extended$ excitations are the
$true$ quanta of a background independent quantum gravity. Thus
in our approach  spacetime becomes a ``process  in the making``
\cite{aa1} emerging from the {\bf C}-space.

It is not difficult to see that the above degeneracy $dg(N)$ is
given by the following expression

\begin{equation}
\label{eq:32}
 dg(N)= \frac{\Gamma(2^D +N)} {\Gamma(N+1)\Gamma({2^D})}
\end{equation}
where $\Gamma$ is the gamma function. We compare $dg(N)$
[Eq.(\ref{eq:32})] with the asymptotic quantum degeneracy of a
massive (super) string state given by Li and Yoneya \cite{l1}:
\begin{equation}
\label{eq:33}
  dg(n)= exp \hspace{2mm}[2\pi \sqrt {n\frac{d_s-2}{6}}\hspace{2mm}]
\end{equation}
where $d_s$ is the string dimension and the string excitation
level number $n>>1$. To this end we equate expression
(\ref{eq:33}) and degeneracy (\ref{eq:32}) of the first ({\it
collective}) excited state ( $N=2^D$) of the $p$-loop.

This move could be justified on physical grounds as follows. One
can consider different frames in the extended relativity: one
frame where an observer sees strings only (with a given
degeneracy) and another frame where the same observer sees
collective excitations of points, strings, membranes,p-loops,
etc. The degeneracy (being a number) should be invariant in any
frame.

Therefore solving the resulting equation $$ dg(N)=  dg(n)$$ with
respect to $\sqrt{n}$ we get
\begin{equation}
\label{eq:34}
 \sqrt{n} = \frac{1}{2\pi}{\sqrt\frac{6}{d_s-2}}\hspace{2mm}{Ln[
 \frac{\Gamma(2^{D+1})}{\Gamma(2^D+1)\Gamma(2^D)}]}
 \end{equation}
 The condition $n >> 1$ implies that $D >> 1$ thus simplifying
(\ref{eq:34}). If, for example, we set $d_s = 26$ ( a bosonic
string) and use the asymptotic representation of the logarithm of
the gamma function for large values of its argument
\begin{center}
$Ln\Gamma(z) =Ln(\sqrt{2\pi})+(z-1/2)Ln(z)- z + O(1/z)$
\end{center}
then from Eq.(\ref{eq:34}) we obtain the following expression for
the respective entropy ( remembering that the entropy is defined
as the $log$ of the degeneracy):
\begin{equation}
\label{eq:35}
 Entropy = Ln [ dg(n)] \sim \sqrt{n} \approx 2^D \sqrt6\hspace{1mm}\frac{ln2}{2\pi} \sim 2^{D-1}\sim N
 \end{equation}
Let us consider a Schwarzschild black hole whose Schwarzschild
radius $R$
\begin{center}
$R \sim (GM)^{1\over (d-3)}$
\end{center}
The black hole mass $M$ and the string length $l_s \sim R$ obey
the Regge relation $$ l^2_s M^2 \sim R^2M^2 = n $$ which implies
that the world sheet area and mass are quantized in Planck units :
$l^2_s = {\sqrt n} \Lambda^2$ and $ M^2 = {\sqrt n} M^2_{Planck} =
{\sqrt n}\Lambda^{-2}$.  Li and Yoneya \cite{l1} obtained the
following expression for the Bekenstein-Hawking entropy of a
Schwarzchild black hole of a radius  $R \sim l_s$: $$ S_{BH} \sim
{ A \over G } \sim { R^{d-2}\over G} \sim { (GM)^{d-2 \over d-3}
\over G} = G^{ {1\over d-3}  } M^{{d-2 \over d-3}} =RM. $$

From the last two equations Li and Yoneya deduced that  the
$(d-2)$- dimensional horizon area in Planck units is proportional
to $S_{BH}$ and $$S_{BH} \sim \sqrt{n}.$$

Now taking into account Eq. (\ref{eq:35}) we obtain
\begin{equation}
\label{eq:36}
 S_{BH} \sim 2^{D-1}
 \end{equation}
where $2^{D-1}$ is approximately equal $N =MR$.  Based on these
results it is tempting to propose that the exact analytical
expression for the entropy associated with the $first~collective$
excitation of the $p$-loop oscillator is given by the logarithm
of the degeneracy :
\begin{equation}
\label{eq:37} S = ln~\{ { \Gamma(2^D + N )\over \Gamma
(N+1)\Gamma(2^D)}\},~~N = 2^D
\end{equation}
where $2^D$ is the degree of Clifford algebra. We plot the
dependence (\ref{eq:37}) in Fig.1 as a graph of $S=S(N)$.

Moreover, to make connection with the entropy of a black hole in
$d$ dimensions, and its relation to the logarithm of the
degeneracy of the highly excited states of a string living in $d$
dimensions, we imposed the following relation (which is in
agreement with the results of \cite{l1}) between $N=2^D$ and
$(d-2)-$ dimensional horizon area of the black hole
\begin{equation}
\label{eq:38}
 N = 2^D =(R/\Lambda)^{(d-2)}= { A_{d-2} \over c_d G_d }, ~~~ G_d
=\Lambda^{ d -2 }
\end{equation}
 where $c_d$ is in general a dimension-dependent coefficient.
For example,  in $d =4$ the coefficient $c_d =4$. In essence, this
coefficient indicates  that a surface area of a two-dimensional
sphere is $4$ times the area of the holographic disk \footnote{
E.Spalucci,Private communication}

The Newton constant in $d$ dimensions is given by $\Lambda^{ d -2
}$ in units $\hbar = c =1$. For black holes, the area $A$ is the
area $transverse$ to the radial and temporal directions, i.e. it
is a $(d-2)$-dimensional area which precisely matches  the power
of the Planck scale entering the definition of the Newton constant
in $d$ dimensions. The number $N$, the degree of the Clifford
algebra in $D$ dimensions, associated with the $p$-loop
oscillator living in $D$ dimensions, is nothing but the number of
$(d-2)$-dimensional horizon area units (bits) of the black hole.
It is seen that both the area and the tension $quantization$
follow quite naturally from the present theory.

This is a rather curious result: the Shannon entropy of a $p$-loop
oscillator in $D$-dimensional space ( for a sufficiently large
$D$), that is the number $N=2^D$ (the degree of the Clifford
algebra in $D$ dimensions) which is equal to the total number of
bits representing $all$ the holographic coordinates , is
proportional to the Bekenstein-Hawking entropy of a Schwarzschild
black hole in $d$ dimensions. The latter is just the number of
($d-2$)- dimensional area bits in Planck units.

A more rigorous study of the connection between Shannon's
information entropy and the quantum-statistical (thermodynamic)
entropy is given by Fujikawa \cite{fuj}. Because light is trapped
inside the black hole its horizon is also an information horizon.

For large values of $N$ and $2^D$ we now take into account in
Stirling's formula applied to Eq.(\ref{eq:37}) the terms with
accuracy $O(1/N)$  and get the following expression for the
entropy:
\begin{equation}
\label{eq:39}
 S = 2N ln ~2 - { 1\over 2} ln~N - { 1\over 2} ln~ 2 - ln~
\sqrt { 2 \pi } - O ( 1/N ).
\end{equation}

In a particular case of $d = 4$ one has then  $ N = A/4G =
A/4\Lambda^2 $, and we will recover not only the logarithmic
corrections recently reviewed in the literature \cite{pm} but
also $all$ the higher order corrections to the Bekenstein-Hawking
entropy in $ d = 4 $ dimensions, up to numerical coefficients:
\begin{equation}
\label{eq:40} S/2ln2 =  {A\over {4 \Lambda^2}} - { 1\over {4 ln
2}}ln{A\over{4\Lambda^2}} - { 1\over 2}[1 + {ln \pi \over{2ln
2}}] - {4\Lambda^2/A\over{2 ln2}}   +...
\end{equation}

As could be expected, Einstein's gravity ( an effective theory) is
recovered in the $long$ range limit, which means that for $N >>1$
a general equation (\ref{eq:37}) yields a linear dependence
between $S$ and $A$. At smaller scales  a departure from this
behavior is observed, since at Planck scales the new (extended)
scale relativistic effects (and associated with it
Cantorian-fractal spacetime, and the respective more general
geometries) begin to take over resulting in deviations from a
linear dependence $S=S(A)$ between entropy and area.

Now we can revisit our previous discussion ( section $2$) of the
temperature of a black hole and determine it with the help of
\begin{equation}
\label{eq:41}
 1/{\bf T} = \partial S/\partial E = ({\partial S
\over
\partial N})({\partial N \over \partial A})({\partial A \over
\partial R})({\partial R \over \partial M})({\partial M \over
\partial E})
\end{equation}
Using  $$A \sim R^{d-2}, R_{bh} \sim (GM)^{1/(d-3)},
G=\Lambda^{d-2}$$ and E =M  (neglecting the back reaction of the
black hole on geometry)  and Eq.({\ref{eq:38}) we find that
\begin{equation}
{1 \over {\bf T}} \sim {1 \over {c_d }}{(d-2) \over(d-3)}(G
M)^{1/ (d-3)}{\partial S \over \partial N}
\end{equation}
We consider the case  of  $~d=4$ which means that $G=\Lambda^2$,
$N \sim \lambda^2M^2 \sim \Lambda^2M^2$, according to (23). Upon
substitution of Eq.({\ref{eq:39}) into Eq.(42) we get: $${ 1
\over {\bf T}}\sim {\Lambda^2 \over 2} (2 M ln2 - {1
\over{2\Lambda^2M}}+ {1 \over {\Lambda^4M^3}}+...)$$ If we
restrict our attention only to the first three terms of this
relation, we find that it reaches its minimum at $M \approx
1.14/\Lambda$. Therefore this will determine the maximum
temperature which can be reached in a black hole: $${\bf T}_{max}
\sim 1/\Lambda \equiv T_{Planck}$$ This is a rather surprising
result telling us that the evaporation of the black hole
(associated with a decrease of its area $A$) is limited by the
upper bound on its temperature, and this upper bound is of the
order of the Planck temperature. In a sense it is not unexpected,
since one of the postulates of the extended relativity is the
existence of absolute scales, Planck scales.

From the entropy-area expression (\ref{eq:40}) we find that the
following condition  must be hold between the entropy of a black
hole of an area $A_3$ (expressed in terms of $N_3$ bits) and the
entropies of two black holes of areas $A_1$ (expressed in terms
of $N_1$ bits ) and $A_2$ (expressed in terms of $N_2$ bits )
whose "fusion" forms a black hole of area $A_3$

$$ (2ln2) (N_3-N_2-N_1) +{1 \over 2} ln \frac{N_1N_2}{N_3}\geq
0.$$ The respective $necessary$ condition looks like a relation
between $N_3$,  a squared mean geometric, and a double arithmetic
values of $N_1$ and $N_2$:

\begin{equation}
\label{eq:42}
 N_1N_2 \geq  N_3 \geq N_1+N_2
\end{equation}

If we consider the entropy of a black hole with an area (in
Planck units) $A_1A_2 \geq A_3 \geq A_1+A_2$ ( where these limits
determine the upper and lower bounds on the entropy of two
"fused" black holes) vs. the sum of the entropies of two black
holes of areas $A_1$ and $A_2$ then using (\ref{eq:37}) we obtain
the curves shown in Fig.2. The graph clearly demonstrates that
the entropy corresponding to $A_3$  is always greater than the
sum of the entropies corresponding to $A_1$ and $A_2$, in full
compliance with the second law of thermodynamics. If one would
define the exponential of the entropy as the $number {\cal N} $ of
micro-physical states then we get:
\begin{equation}
\label{eq:43}
 {\cal N}  = e^S = e^{ 2N ln2 - { 1\over 2} lnN -
...} \sim 2^{2N}N^{-1/2} = { 2^{2^{D+1}} \over 2^{D/2}}
\end{equation}
Notice the {\bf double exponents} ( a "googolplexus") appearing in
the number of micro-physical states .

For a $k$-th excited level  of the $p$-loop oscillator we obtain
the following expression:
\begin{equation}
\label{eq:44}
 N_k = k N_1 =k {(R/ \Lambda)}^{d-2}= k 2^D = { k A_{d-2} \over c_d G_d }
= { k A_{d-2}  \over c_d \Lambda^{d-2} }
\end{equation}
where we use the exact relation $N={(R/ \Lambda)}^{d-2}$.  The
number $N_k$ may correspond to the entropy of a bound state of
many black holes.

\section{Conclusion}
Application of  a simplified linearized equation derived from the
master action functional of an extended relativity to a $p$-loop
oscillator has allowed us to elementary obtain rather interesting
results. First of all, the solution explicitly indicates
existence of $2$ extreme regions characterized by the values of
the dimensionless combination $m_{p+1}\Lambda^{p+1}$ :

\smallskip
 $1)$ the fractal region where  $m_{p+1}\Lambda^{p+1}\sim
1$ and $2$ scales collapse to one, namely Planck scale $\Lambda$

and

$2)$ the smooth region where $m_{p+1}\Lambda^{p+1}<< 1$ and we
recover the description of the conventional quantum point
oscillator.

Here $2$ scales are present , a characteristic "length" $a$  that
we identified with the string scale $l_s$ and the ubiquitous
Planck scale $\Lambda$ ( $a > \Lambda$) thus demonstrating
explicitly the implied validity of the quantum mechanical
solution in the region where $a/\Lambda > 1$.

For a specific case of $p=1$ (a string) the solution yields (once
again in an elementary fashion) a quantization of the sting
tension $T$ and one of the basic relations of string theory $T =
1/\alpha'$. In addition, we derive a string uncertainty relation
which in turn is a truncated version of a more general string
uncertainty relation derived earlier \cite{cc4},\cite{cc5}.

A simple analysis  based on the found quantization of the p-loop
oscillator tension allowed us to derive both the Schwarschild
radius-mass relation for a black hole in $d$ dimensions and the
generalizations of the famous "Hawking temperature" $\bf T$ for an
arbitrary number of dimensions (not necessarily $4$): $${\bf T}
\sim(GM)^{1/(3-d)}.$$

Moreover, we evaluated the upper bound on the  temperature
achieved in a black hole. It has turned out that this bound is of
the order of the Planck temperature. Since a decrease of the black
hole area is accompanied by an increase of its temperature, this
indicates that the black hole evaporation cannot continue
"unchecked" and stops when the temperature reaches the Planck
temperature.

We equated the degeneracy of the $first~collective$ state of the
$p$-loop oscillator to  the degeneracy  of highly excited state of
the massive ( super) string theory, given by Li and Yoneya
\cite{l1}. As a result, it is found that the Shannon entropy of a
$p$-loop oscillator in $D$-dimensional space (where $N=2^D$ is
the number of bits representing $all$ the holographic
coordinates) is proportional to the Bekenstein-Hawking entropy of
the Schwarzschild black hole in $d$ dimensions. In the linear
regime of entropy versus area, the latter is just the number of
$( d-2 )$- dimensional area bits in Planck units.

Moreover, the solution allowed us to find (up to numerical
coefficients) the $logarithmic$, and higher order corrections to
the entropy-area linear law  (in full agreement with \cite{pm})
by retaining the first order terms in Stirling's formula. It is
also suggestive to propose  that the exact analytical expression
for the black hole entropy is associated with the
$first~collective$ excitation of the p-loop oscillator and is
given by Eq.(\ref{eq:30}) $$S =Ln\{
\frac{\Gamma(2^D+N)}{\Gamma(N+1)\Gamma(2^D)}\},~ N =2^D=
\frac{A_{d-2}}{c_d G_d},~ G_d =\Lambda^{d-2}$$ where $c_d$ is a
dimension-dependent coefficient which for $d=4$ is $c_4 = 4$.

Thus a study of a simplified model ( or "toy") problem of a
linearized $p$-loop oscillator based on importance of collective
modes of excitations ( in compliance with the recent results by
Chapline, Laughlin {\it et. al} with regards to the black holes)
gives us ( with the help of elementary calculations) not only a
set of the well-known relations of string theory but also the
logarithmic correction to the well-known black hole entropy-area
relation ( obtained earlier on the basis of a much more
complicated mathematical technique). This indicates that the
approach advocated by the extended relativity might be very
fruitful, especially if it will be possible to obtain analogous
analytic results on the basis of the full master action
functional and the resulting functional nonlinear equations whose
study will involve non-trivial extensions of the Clifford
algebra, e.g. the braided Hopf quantum Clifford algebras.
\bigskip
\newpage
\centerline{\bf Acknowledgements }

The authors would like to thank E.Spallucci and S.Ansoldi for
many valuable discussions and comments. Special thanks to A.
Quiroz for his valuable comments.

\end{document}